\documentclass[conference]{IEEEtran}  

\usepackage{graphics} 
\usepackage{amsmath} 
\usepackage{amssymb}  
\usepackage{graphicx}
\usepackage{subfig}
\usepackage{array}
\usepackage{cite}
\usepackage{cellspace}
\usepackage{authblk}
\newcommand\cincludegraphics[2][]{\raisebox{-0.2\height}{\includegraphics[#1]{#2}}}

\makeatletter
\def\endthebibliography{%
  \def\@noitemerr{\@latex@warning{Empty `thebibliography' environment}}%
  \endlist
}
\makeatother

\begin{document}
\title{
Substation One-Line Diagram Automatic Generation and Visualization
}

\author[a]{Jing Hong}
\author[b]{Yue Li}
\author[c]{Yiran Xu}
\author[a]{Chen Yuan}
\author[a]{Hong Fan}
\author[a]{Guangyi Liu}
\author[a]{Renchang Dai}

\affil[a]{GEIRI North America, San Jose, CA, USA\thanks{This work is supported by the State Grid Corporation technology project 5455HJ180020.}}
\affil[b]{College of William \& Mary, Williamsburg, VA,  USA}
\affil[c]{State Grid Nanjing Power Supply Company, Nanjing, Jiangsu, China}



\maketitle

\begin{abstract}

In Energy Management System (EMS) applications and many other off-line planning and study tools, one-line diagram (OLND) of the whole system and stations is a straightforward view for planners and  operators to design, monitor, analyze, and control the power system. Large-scale power system OLND is usually manually developed and maintained. The work is tedious, time-consuming and ease to make mistake. Meanwhile, the manually created diagrams are hard to be shared among the on-line and off-line systems. To save the time and efforts to draw and maintain OLNDs, and provide the capability to share the OLNDs, a tool to automatically develop substation based upon Common Information Model (CIM) standard is needed. Currently, there is no standard rule to draw the substation OLND. Besides, the substation layouts can be altered from the typical formats in textbooks based on factors of economy, efficiency, engineering practice, etc. This paper presents a tool on substation OLND automatic generation and visualization. This tool takes the substation CIM/E model as input, then automatically computes the coordinates of all components and generates the substation OLND based on its components attributes and connectivity relations. Evaluation of the proposed approach is presented using a real provincial power system. Over 95\% of substation OLNDs are decently presented and the rest are corner cases, needing extra effort to do specific reconfiguration. \\
\end{abstract}

\IEEEpeerreviewmaketitle

\section{Introduction}
With high penetrations of renewable energy resources, distributed generation, and nonlinear power electronics devices, the wellness of system operation is facing more challenges \cite{textbook, 8274661, 7356784, mao2019integrated}. To help system operators monitor and supervise power grids operation, automatic generation and visualization of the substation one-line diagram (OLND) is necessarily needed \cite{4498697, 5647311, 8586535}. 

In the existing Energy Management System (EMS), few substation OLND layouts are automatically developed. Most of them are manually created based on a special drawing tool and specific OLND creation approach \cite{8408545}. But, it has great potential to cause the inconsistency between database and visualized substation OLNDs, since changes made on either back-end database or the front-end substation OLNDs might not be automatically synchronized. It is even more probable to happen in developing countries, since power grids in developing countries are experiencing rapid development and having numerous substations being built and expanded \cite{yachendiss}. Currently, there are three main reasons impeding the automation of substation OLND drawing: (1) diverse substation configurations; (2) various connections between substation components; (3) complicated components placement. Regarding the first factor, the real substation OLNDs are different from the classical substation layouts and diversified with compromise of economy, efficiency, engineering practice, etc. Meanwhile, the classical substation OLNDs are extracted from the real cases by keeping the common configurations and trimming different parts. On the other hand, the voltage level and the number of voltage levels could be very different in each substation. For example, the configuration of a substation with three-winding transformer(s) is totally different from a substation without transformer or with a two-winding transformer. The second barrier is the variability or flexibility of components connections. Taking disconnector as an example, a disconnector could have two breakers connected on its both sides, or have one breaker and one bus/load/generation unit/transformer/compensator/AC line connected with it, or have a T-type connection to connect another disconnector and a circuit breaker on the other side, etc. When it comes to the third factor, the components positions are not fixed and can be randomly assigned in the drawing of substation OLNDs, since there is no standard rule to place and layout components. However, to automatically present a decent substation OLND layout, each component needs to coordinate with others to avoid component overlap, minimize line crossings, and get evenly distributed \cite{699362}.

While automatic generation algorithms have been developed for OLNDs in both distribution and transmission networks \cite{8408545, 699362, 4523781, Zhao2017GraphbasedPC}, previous research has shown limited success in automatic generation of substation OLNDs. A graphics model in common information model (CIM) framework expressed in unified modeling language (UML) was introduced in 2010 to define semantics of the graphical information and associate the graphic object to corresponding domain data object in CIM. Similarly, CIM based graphic exchange format (CIM/G) was proposed in 2011 to extend the standard CIM model to include the schematic layout information \cite{5772511}. A graphics model using layout and relative coordinates was proposed to extend the initial CIM graphics UML model while considering use cases including relative coordinates, layout mechanism and auto-generation \cite{6523222}.

In this paper, a substation OLND automatic generation and visualization approach is demonstrated. This substation OLND drawing tool takes the substation CIM/E model as input, then automatically computes the coordinates of all components and well generates the substation OLND based on its components attributes and connectivity relations. It provides system operators a vivid picture of the substation topology and has the capability of displaying the static information and dynamic status of the system. 


In Table~\ref{tb:comp}, eight types of components usually installed in substations are listed. Since there are no unified symbols for these components, the ones we used might not be exactly the same as other models'. However, this is not a problem as the symbols can be easily added, modified or replaced in the presented tool. Furthermore, the contribution of the paper is mainly demonstrating the automatic generation of the substation OLND layout with intelligent assignment of components coordinates.



The paper is organized as follows. In Section~\ref{methodology}, The proposed substation OLND generation and visualization approach is fully elaborated. The evaluation of the approach and discussion of its pros and cons are presented in Section~\ref{discussion}. At last, the paper is concluded in Section~\ref{conclusion}.

\renewcommand{\arraystretch}{1.7}
\begin{table}[h]
  \centering
  \caption{Components in the Substation Topology}
\begin{tabular}{| >{\arraybackslash}m{1.5in} | >{\arraybackslash}m{1in} |}
\hline
Name                   & Symbol \\ \hline
Breaker                & 
\cincludegraphics[height=4mm]{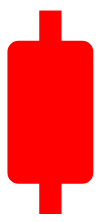}
\cincludegraphics[height=4mm]{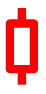}
\\ \hline
Disconnector           &
\cincludegraphics[height=4mm]{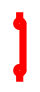}
\cincludegraphics[height=4mm]{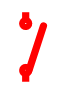}
\\ \hline
AC line                &
\cincludegraphics[height=4mm]{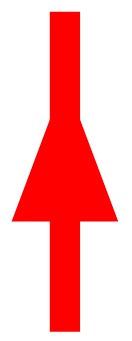}
\\ \hline
Load                   &
\cincludegraphics[height=4mm]{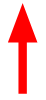}
\\ \hline
Three-winding Transformer &
\cincludegraphics[height=5mm]{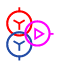}
\\ \hline
Two-winding Transformer   &
\cincludegraphics[height=5mm]{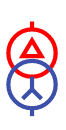}
\\ \hline
Compensator            &
\cincludegraphics[height=4mm]{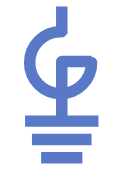}
\\ \hline
Generation Unit        &
\cincludegraphics[height=4mm]{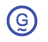}
\\ \hline

\end{tabular}
\label{tb:comp}
\end{table}

\section{Methodology}
\label{methodology}
\subsection{Overview}

\begin{figure}[h]
\centering
\includegraphics[width=.48\textwidth]{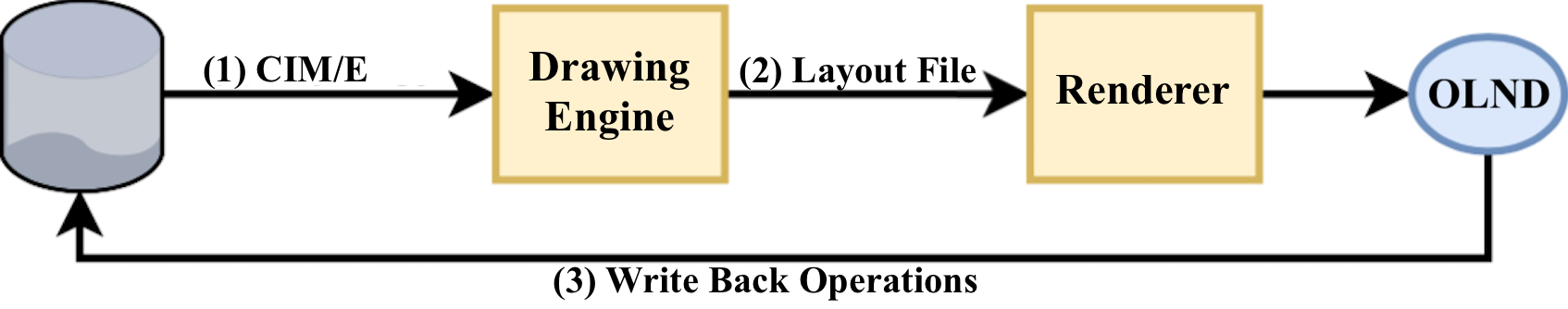}
\caption{An overview of the substation OLND automatic generation and visualization tool}
    \label{fig:overview}
    \centering
\end{figure}

Fig.~\ref{fig:overview} presents an overview of the proposed substation OLND generation and visualization tool. The input is a CIM/E model representing the components in an electrical substation and their connectivity relations \cite{zhangxin}. CIM/E is an efficient electric grid model exchange standard format developed by State Grid Corporation of China (SGCC). The CIM/E model simplifies CIM/XML model by ignoring terminal without compromising information to
process network topology. In this paper, substation CIM/E models are stored in a graph database and they can be queried by substation name. With the input information, a substation OLND is automatically generated and its visualization is the output. 


This tool consists of two phases, (1) drawing and (2) rendering.
In the drawing phase, a drawing engine is developed to query the CIM/E data from the graph database, and output a Json-style file that specifies the symbols, positions, orientations, and status of electrical components, as well as line configurations connecting these components. In the rendering phase, the renderer takes in the Json file that is specifically tailored for it, and interprets the information. The actual rendering of the OLND is then implemented on a web page by the renderer. 

In the following, the detailed implementation of each phase is introduced.

\subsection{Drawing}
\begin{figure}[h]
\centering
\includegraphics[width=.48\textwidth]{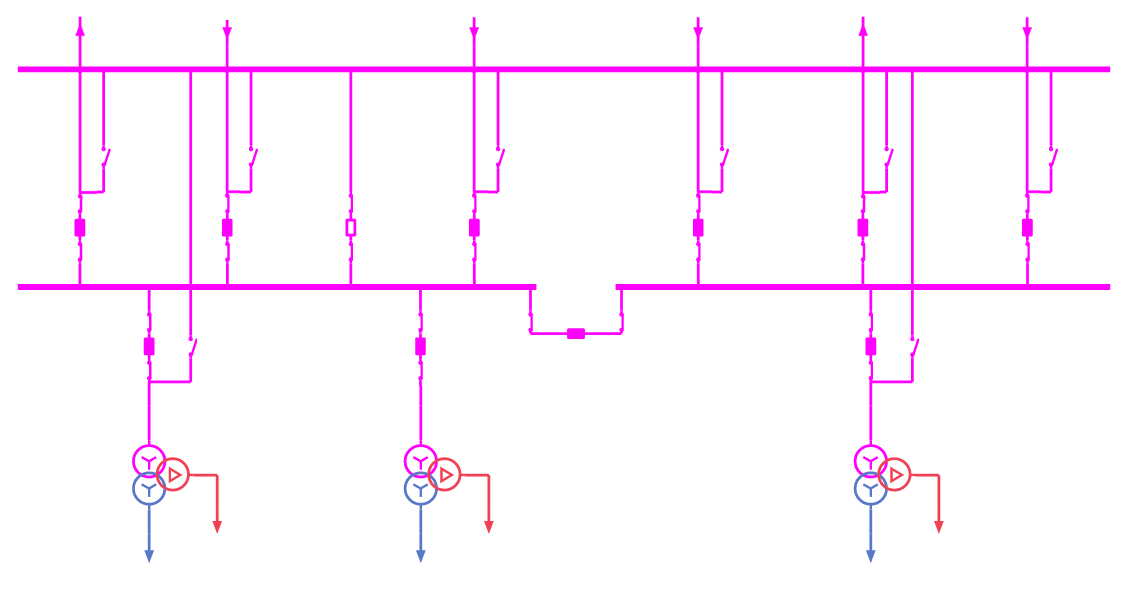}
\caption{An example of a drawn substation OLND}
    \label{fig:example}
    \centering
\end{figure}

The most challenging task in the drawing phase is to determine the appropriate layout scheme of the substation OLND, and compute the position of each component in the substation OLND. The drawing engine is in charge of this task and it can be naturally decomposed into several layers. In each layer, the positions of a specific collection of components are determined. For instance, in the first layer, the bus positions are determined, and then the positions of branches connected to the buses are relatively located. A branch may have one or more sub-branches, and each sub-branch is like a path in a graph. The sub-branches positions are calculated in the next layer, and the coordinate of each component connected to sub-branches can be assigned. An example of a substation OLND is visualized in Fig.~\ref{fig:example}. This substation OLND drawing is quite representative and well reflects the design thoughts of the drawing engine.

The input CIM/E data from the graph database include nodes, edges and their attributes. The edges, which have directions, are stored in a HashMap, where the key is the from node ID and the value is an array of its neighbors' node IDs. Besides, another map is also maintained to store the attributes of each node. These attributes include the voltage level, name, status, etc. of each component. The output of the drawing engine is a Json-styled file that specifies necessary information to draw the components. For example, the Json representation of a disconnector is like the following.
{\small
\begin{verbatim}
{
  "p": {
    "position": {
      "y": -281.0, 
      "x": -458.0
    }, 
    "tag": "Disconnector#282", 
    "image": "symbols/disconnector.json"
  }, 
  "c": "Node", 
  "a": {
    "state": true, 
    "voltage": "500kV", 
    "lineColor": "rgb(255,0,0)", 
  }
}
\end{verbatim}
}

The substation OLND is composed by a collection of component representations, and the actual drawing will be done by the drawing engine.

\subsubsection{Layout of Different Voltage Levels}

The first step is to find out the number of voltage levels, and group them together. The voltage levels are easily retrieved from the attributes of bus nodes in the graph database. For those voltage levels only associated to transformers instead of buses, as they do not have an impact on the positioning of buses, they are temporarily ignored at this step. 

\begin{figure}[h]
\centering
\includegraphics[width=.48\textwidth]{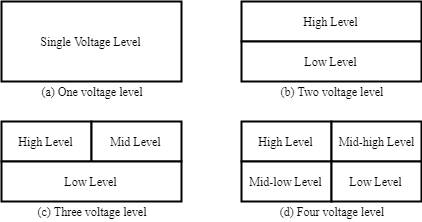}
\caption{Layouts of different voltage levels}
    \label{fig:voltage_layout}
    \centering
\end{figure}

The buses with different voltage levels are drawn together and occupy their own regions, named as \textbf{voltage region}, in the OLND. The area size of each voltage region is determined by the bus scheme, bus size, and the components that connect to the bus. We employ the pattern of $(maxX, minX, maxY, minY)$ to record the size of the voltage region, which is in the form of a rectangle, and arrange the voltage regions according to the their sizes. 
The relative placement of different voltage regions depends on the magnitude of the voltage level. The detailed layout is shown in Fig.~\ref{fig:voltage_layout}. As shown in the figure, the canvas splits into an upper level and a lower level when there exist more than one voltage regions. These regions are usually connected via two-winding or three-winding transformers. Note that the voltage region layout identified at this step does not always apply. There are a few cases' OLND layout should be modified according to the positions of two-winding transformers, which is elaborated in Section~\ref{sec:transformer}.

\subsubsection{Layout of Buses}
\label{sec:buslayout}
\begin{figure}[h]
\centering
\begin{tabular}{cc}
\subfloat[\small Double bus single breaker] { \includegraphics[width=0.45\linewidth]{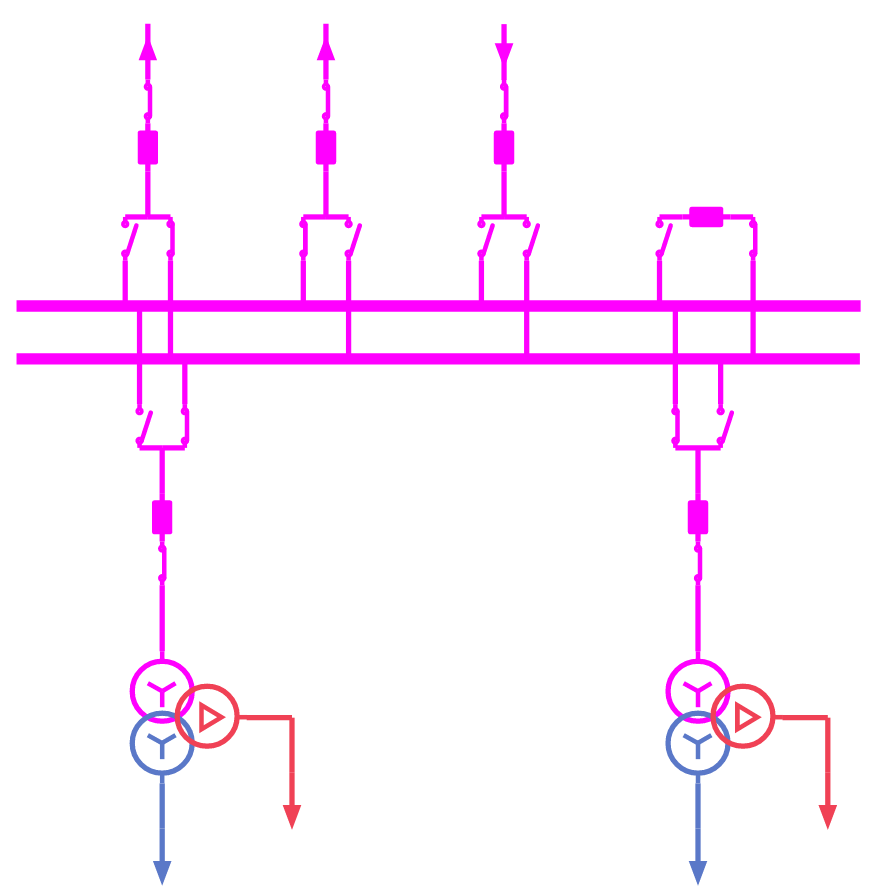} \label{fig:singlebreaker}} 
&
\subfloat[\small Main bus and bypass bus] { \includegraphics[width=0.45\linewidth]{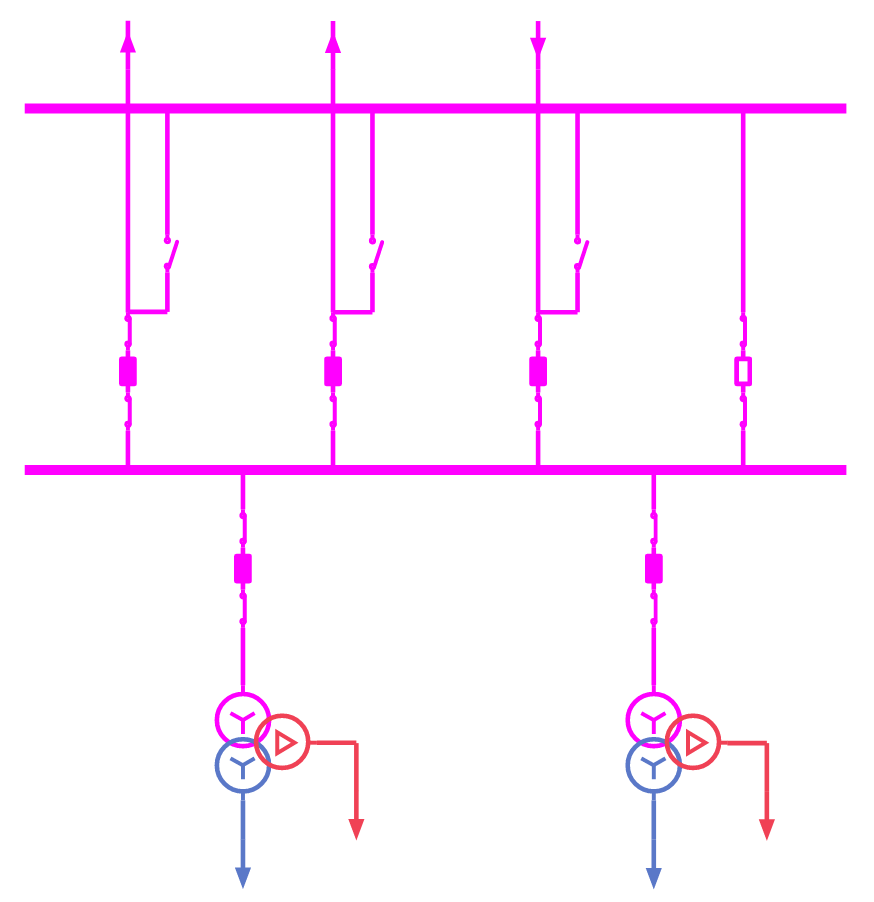} \label{fig:transferbus}}

\\
\subfloat[\small Breaker and half / double bus double breaker] { \includegraphics[width=0.45\linewidth]{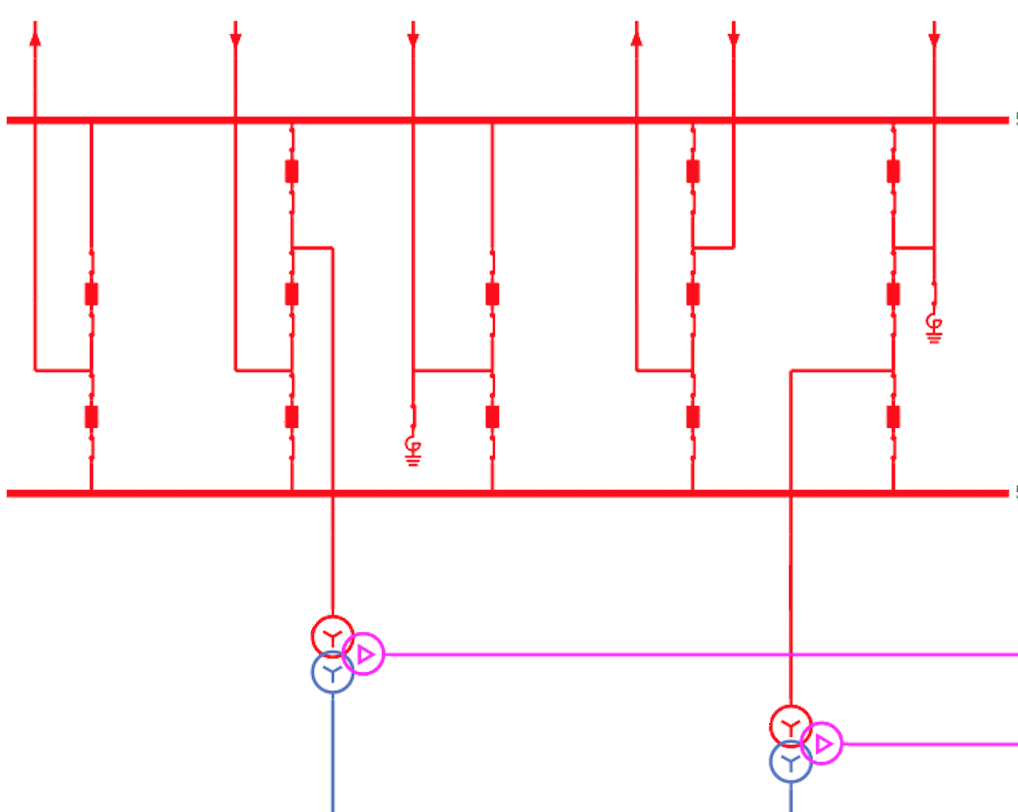} \label{fig:32}}
&
\subfloat[\small Single bus with sectionalizer] { \includegraphics[width=0.45\linewidth]{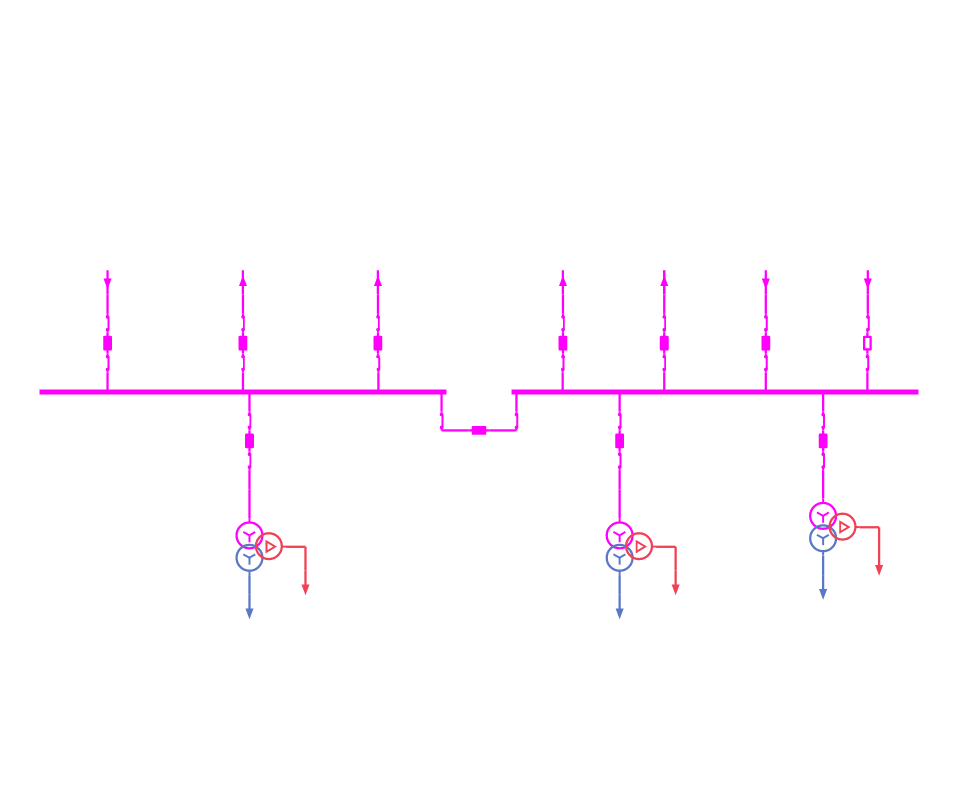} \label{fig:sectionalizer}}
\\
\end{tabular}
\caption{Bus schemes of two buses \cite{123456}} 
\label{fig:schemes}
\end{figure}

Within a substation, each voltage level may have multiple buses to follow common bus scheme. In other words, the bus scheme is determined for all the components in the same voltage level of the substation. Then, each bus position is allocated. For example, Fig.~\ref{fig:schemes} illustrates OLNDs for all the two-bus schemes that we have observed from our database, and each OLND represents a physical substation layout \cite{123456}. Besides, it is needed to note that, in the graph database, a single bus with a sectionalizer is defined as two buses, which is also depicted in Fig.~\ref{fig:sectionalizer}). Besides, the \textit{Double Bus Double Breaker} bus scheme and the \textit{Breaker and Half} bus scheme resembles each other in the way of drawing and the only difference is the number of circuit breakers and disconnectors (as shown in Fig.~\ref{fig:32}) . As such, they are grouped and drawn using the same program routine.

In total, the system has more than 10 bus schemes and each has one to four buses in the voltage region. Due to space limit, this paper only describes how the four schemes in Fig.~\ref{fig:schemes} are determined and positioned. Other schemes follow a similar approach. 

Essentially, the bus scheme is identified by the connections between buses. Each component that is directly connected to the bus is called as a \textbf{branch head}, and the collection of all components that is directly or indirectly connected to the branch head without passing a bus or a transformer, is called as a \textbf{branch}. Note that the components in a branch may be shared by two buses if a sub-branch connects to the two buses. 

The first step is to find all branches connected to each bus, which can be easily done through a depth-first traversal on the HashMap representation of the OLND. Some branches have strong indications of the bus scheme. For example, in Fig.~\ref{fig:singlebreaker}, we know it is a double bus scheme as the two buses are connected by many $\{Disconnector, Disconnector\}$ sub-branches. On the other hand, if each bus has many branches that are not related to the other bus (except for a $\{Disconnector, Breaker, Disconnector\}$ sub-branch as the bus connector), the two buses are determined to be the ``Single Bus with Sectionalizer" pattern, as shown in Fig.~\ref{fig:sectionalizer}. Similarly, bypass bus can be identified by examining if there are multiple $\{Disconnector, Breaker, Disconnector, Disconnector\}$ sub-branches, and the bus with two consecutive disconnectors are the bypass bus. Finally, the ``Breaker and Half" and the ``Double Bus Double Breaker" schemes are treated similarly. They are identified by a sub-branch connecting two buses with two or more $\{Disconnector, Breaker, Disconnector\}$ segments. 

After the bus scheme is figured out, the length of the bus will be calculated. As the bus is represented by a line in the substation OLND, the length of the line is determined by the number of branches connected to the bus and the width of each branch. We compute the actual width of each branch, and add a pre-set distance between branches. The sum of them determines the total length of the bus. Note that the bus may have branches going upward or downward in the substation OLND. The compuation should be separate and the larger one is used to set the length of the bus. The length computation is done by running the same logic of drawing branches but without including the branch data in the output Json file. 

\subsubsection{Transformers}
\label{sec:transformer}

\begin{figure*}[h]
\centering
\includegraphics[width=1\textwidth]{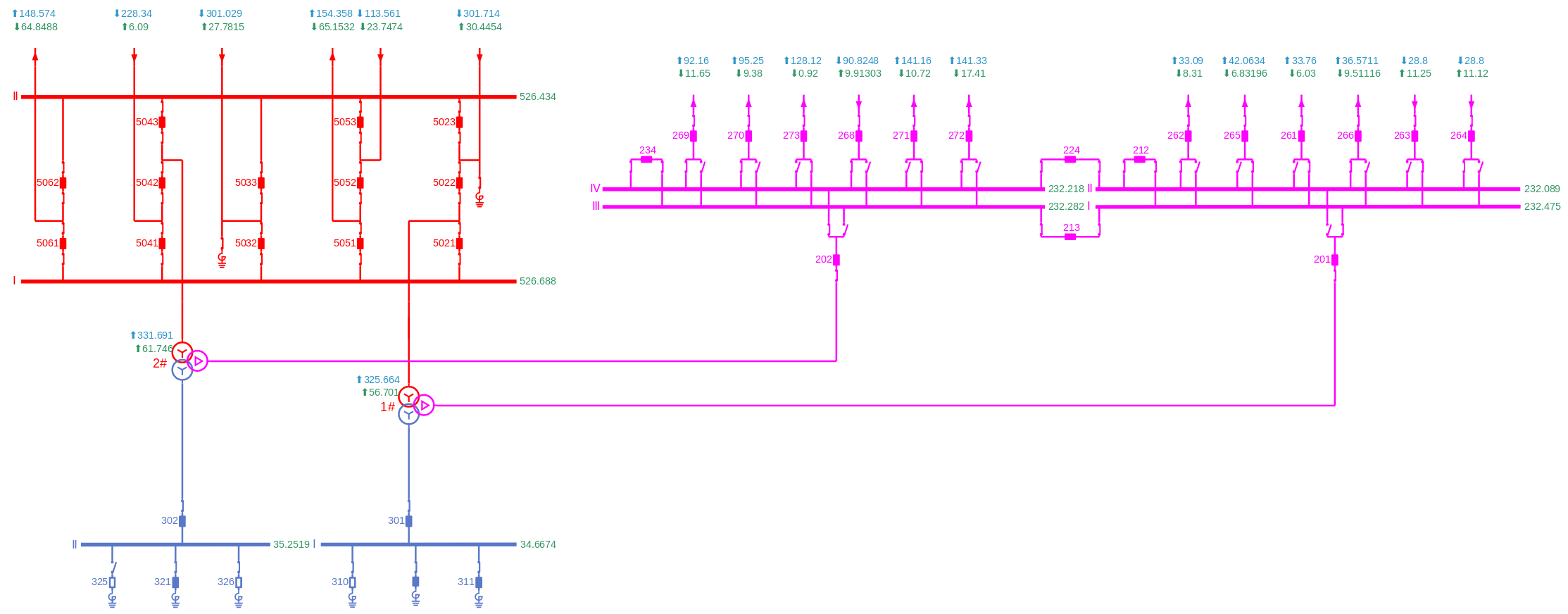}
\caption{Example of a generated substation OLND}
    \label{fig:final}
    \centering
\end{figure*}

The transformer is special as it is a component that connects other components in different voltage levels. For the implementation, transformers are also critical and may impact positions of others. This is because a transformer may decide the positions of two branches. For example, in Fig~\ref{fig:final}, the two three-winding transformers decided the positions of the two blue branches right below them. 

\subsubsection{Branches}
\label{sec:branches}
As described in Section~\ref{sec:buslayout}, branches are defined as all components that are directly or indirectly connected to a bus. After the position and length of a bus are determined, we are able to draw branches connected to it. The first step is to identify directions of these branches. By default, all branches are placed above their connected buses and direct upwards, but there exist some exceptions. If one terminal of the branch is connected to a generator, it is supposed to be placed below its connected bus and point downwards. If there exists any transformer in the branch, the direction should be determined by its voltage level relative to other branches connected to the same transformer. For example, if the branch being drawn is at the lower voltage level than the other branch connected to the same two-winding transformer, or at the lowest voltage level compared to other branches connected to the same three-winding transformer, its direction should remain unchanged. Otherwise, we reverse its direction to place the branch below its connected bus, so that the transformer can always be placed in the middle to avoid potential line crossings and component overlaps. 

After the directions of all branches are determined, their positions relative to each other on the bus will be fixed. Though a branch may be connected to multiple buses, only one bus is in charge of drawing this branch. Branches placed above the buses are sorted by the number of nodes in branches, and check if they are connected to another vertically or horizontally paired bus. For branches placed below the buses, they are sorted based on their positions connected to the transformers. After all branches are sorted, their positions are finalized. The default gap between every two branches are determined by the number of branches in the same direction. However, it is possible that some branches have more sub-branches than others, and therefore it needs more space to avoid component overlap on canvas. Thus, it is needed to estimate the width of each branch in advance, and assign extra space when needed. All sub-branches are found and their widths are computed through a depth-first traversal on the HashMap representation of the substation OLND. 

\begin{figure}[h]
\centering
\includegraphics[width=.4\textwidth]{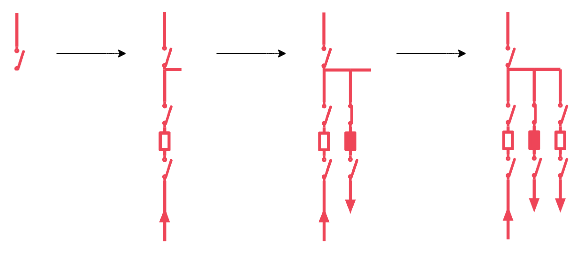}
\caption{Process of drawing a branch}
    \label{fig:subbranchs}
    \centering
\end{figure}

After all branches positions are fixed, their sub-branches are drawn. For each branch, every sub-branch it contains will be drawn recursively. Taking the branch in Fig~\ref{fig:subbranchs} as an example, starting from the $Disconnector$, it could be found that it has multiple sub-branches. We assign different directions to these sub-branches. Only one sub-branch is in the same direction as the branch, and the remaining are supposed to be placed to its right. After the directions of sub-branches are determined, the one following its direction is drawn directly. For each sub-branch placed right, we apply the same method to find their connected sub-branches and assign them directions. When drawing the first sub-branch in the right direction, more than one sub-branches are found. One in the same direction with the branch is placed, and others to the right. This procedure will be done recursively, until all sub-branches have been drawn.

\subsection{Rendering}
In the rendering phase, the render takes in the Json file generated by the drawing engine, which specifies the symbol, position, status, tag, and other information of every component and AC transmission line. Then, it interprets and visualizes all information using Hightopo JavaScript UI library. 



\section{Discussion}
\label{discussion}
The performance of the proposed substation OLND automatic generation and visualization tool is evaluated over 799 substations in a real provincial power grid. We successfully generate the OLNDs for all substations, and over 95\% of them are decently presented. Fig.~\ref{fig:final} is an example of a substation OLND generated by the developed tool. It includes three voltage levels, i.e. 500 kV, 220 kV, and 35 kV. In this example, it employs two bus schemes,\textit{ Breaker and Half} and \textit{Double Bus Double Breaker}, as displayed in Fig.~\ref{fig:32}, for 500 kV part, and \textit{Double Bus Single Breaker} bus scheme for 220 kV level, as presented in Fig.~\ref{fig:singlebreaker}. All three voltage levels are connected via three-port transformers, as shown in Fig.~\ref{fig:final}. But, there still exists some unsolved patterns, such as overlapped lines and components, in 34 substations of the total. Because the position of each component could be influenced by other components and it is very difficult to cover all corner cases without manual update of certain component positions. For those OLNDs with unexpected patterns, it needs extra efforts to specifically improve their configurations. Besides, since the coordinates of all components are calculated based on the attributes and connections of the input CIM/E model, this tool is hardly to generate a decent substation OLND if the topology contains any data that violates the basic assumptions. For example, it is currently assumed that the input data only contains one to four voltage levels associated with buses, so the tool is unable to find an appropriate layout if it contains five voltage levels. But in the future work, this could be figured out and becomes more intelligent.

\section{Conclusion}
\label{conclusion}

In this paper, a tool on substation one-line diagram automatic generation and visualization is presented. With the input of CIM/E model, which is stored in a graph database and displays component connectivity in the substation, the proposed approach is able to intelligently compute the positions of all components and decently visualize substation one-line diagrams. Over  95\%  of substation one-line diagrams in a real provincial grid are decently presented in the testing.




\bibliography{IEEEabrv,IEEEexample}
\bibliographystyle{IEEEtran}

\end{document}